# Partial Template Based Receiver in Impulse Radio Ultra-Wideband Communication Systems

Cao Yang
School of Electronics and Information Engineering
Harbin Institute of Technology
Harbin, China
xiaoyang_cao@126.com

Luo Chao
School of Electronics and Information Engineering
Harbin Institute of Technology
Harbin, China
chluochao@163.com

Wu Xuanli
School of Computer Science and Technology
School of Electronics and Information Engineering
Harbin Institute of Technology
Harbin, China
xlwu2002@hit.edu.cn

*Abstract*—For high speed ultra-wideband (UWB) communication systems, the multipath interference exhibits a primary obstacle to improve transmission performance. In order to enhance the signal to interference plus noise ratio (SINR) in the receiver, a partial template receiver is proposed in this paper. Instead of using the conventional template, the model in this paper adopts a partial template to demodulate signals. To analyze the performance of the proposed receiver, bit error rate (BER) formulation of IR-UWB systems in the presence of multipath interference, multiuser interference and addictive white Gaussian noise (AWGN) is derived in IEEE 802.15.4a channel models. Simulation results show that, compared with the conventional correlation receiver, the proposed receiver can achieve a better BER performance for high $E_b/N_0$, which falls in the conventional used $E_b/N_0$ range.

## I. INTRODUCTION

As an energy efficient and low-complexity wireless access technology, ultra-wideband (UWB) has attracted more and more researchers' attention in wireless sensor networks (WSN) and wireless body area network (WBAN) [1-3]. UWB is selected as one of the enabling technologies for wireless sensor networks in 2007 [1-2]. However, in certain specific environments, especially residential and indoor office environments, the dense multipath channel of UWB becomes a big challenge to collect energy efficiently with huge multipath interference. Generally, there are two types of multipath interferences: One is caused by the interference of two adjacent data symbols, which is called inter-symbol interference (ISI). Another is caused by the interference between a pulse and its own multipaths, which is called intra-symbol interference (IASI). And both of them will have a significant impact on the BER performance of UWB systems, especially when the data transmission rate is high. In recent years, many researchers have done a lot of research on waveform designing to maximize power efficiency of pulse restricted by FCC spectral mask. Undoubtedly, an effective use of power efficiency can improve SNR in the receiver; however, it cannot ensure the receiver can also obtain a high SINR, which includes both noise and interference. Therefore, higher power efficiency cannot guarantee a better BER performance. And [4] and [5] discussed the system transmission performance with different waveforms and finally they drew the conclusion that a good correlation property of waveform should also be taken into consideration instead of only focusing on maximizing power efficiency of pulse. On the other hand, in order to make an effective use of multipath components, different types of receivers are proposed [6-8]. [9] analyzed the performance of RAKE reception and compared two combining strategies-equal gain combining (EGC) and maximal ratio combining (MRC). In terms of hardware complexity, TR receiver has a obvious advantage over RAKE receiver. However, [10] pointed out that the noise can severely decrease the performance of TR receiver because reference pulse must be transmitted in noise channel.

However, simulation in this paper shows that correlation receiver cannot achieve a desirable BER performance in presence of addictive white Gaussian noise, multipath interference and multiuser interference. Therefore, to enhance the performance of correlation receiver, a novel receiver algorithm-partial template receiver is introduced and a detailed analysis of how it can improve system performance is presented.

## II. SYSTEM MODEL OF ULTRA-WIDEBAND COMMUNICATIONS

In a TH-BPSK UWB system, the transmitted signal of the *n*-th user can be given by:

$$s^{(n,i)}(t) = \sum_{j=0}^{N_s-1} d^{(n,i)} \sqrt{E_p} p(t - jT_f - C_{i,j}^{(n)} T_c). \quad (1)$$

where, $p(t)$ is the unit energy pulse waveform, whose energy is $E_p$ and mean pulse repetition period is $T_f$. $C_{i,j}^{(n)} = \{1, 2, 3, \ldots, N_h\}$ is the time hopping sequence of the *i*-th bit of the *n*-th user with $T_c$ corresponding to the time hopping slot time. $d^{(n,i)} \in \{-1, 1\}$ represents the binary data sequence and one data symbol is composed of $N_s$ pulses.

This work is supported by Specialized Research Fund for the Doctoral Program of Higher Education (New Teachers) (Grant No. 20092302120001), China Postdoctoral Science Foundation (Grant No. 20100471080), and Heilongjiang Postdoctoral Grant (Grant No. LBH-Z09153).

According IEEE 802.15.4a, the impulse response of UWB system can be written as follows:

$$h(t) = \sum_{l=1}^{L} \sum_{k=1}^{K} \alpha_{k,l} \delta(t - T_l - \tau_{k,l}). \quad (2)$$

where, $\alpha_{k,l}$ corresponds to the tap weight of $k$-th ray of $l$-th cluster, and $\tau_{k,l}$ is the $k$-th ray's arrival time relative to the $l$-th cluster's arrival time $T_l$. $L$ and $K$ denote the number of observable clusters and rays, respectively. Finally the received signal can be demonstrated as:

$$r(t) = \sum_{n=1}^{N_u} \sum_{i=1}^{N_I} s^{(n,i)}(t - \tau^{(n)} + iT_f) * h(t) + n(t). \quad (3)$$

where, $\tau^{(n)}$ is the $n$-th user's delay relative to the first user due to asynchronous transmission, assuming $\tau^{(1)} = 0$. $N_I$ represents the number of interfering pulses from the previous periods, $N_I = \lceil \tau_{\max} / T_f \rceil = \lceil \tau_{\max} R_b N_s \rceil$, $R_b$ is transmission bit rate, $\tau_{\max}$ is maximum multipath delay. And $n(t)$ denotes addictive white Gaussian noise.

For the tap weight $\alpha_{k,l}$, it follows a Nakagami-m distribution with parameters ($\Omega, m$) according to

$$pdf(\alpha_{k,l}) = \frac{2}{\Gamma(m_{k,l})} \left(\frac{m_{k,l}}{\Omega_{k,l}}\right)^{m_{k,l}} \alpha_{k,l}^{2m_{k,l}-1} \exp\left(-\frac{m_{k,l}}{\Omega_{k,l}} \alpha_{k,l}^2\right). \quad (4)$$

where, $\Gamma(\cdot)$ represents the Gamma function, $m$ is the Nakagami-m factor which is modeled as a lognormally distribution random variable, and $E[\alpha_{k,l}^2] = \Omega_{k,l}$.

The mean power of different rays is given by

$$E[\alpha_{k,l} \alpha_{k_1,l_1}] = \begin{cases} \frac{\Omega_l \exp(-\tau_{k,l}/\gamma_l)}{\gamma_l[(1-\beta)\lambda_1 + \beta\lambda_2 + 1]} & k=k_1 \text{ and } l=l_1 \\ 0 & k \neq k_1 \text{ or } l \neq l_1 \end{cases}. \quad (5)$$

where, $\Omega_l$ corresponds to the integrated energy of the $l$-th cluster, and $\gamma_l$ is the intra-cluster decay time constant, which is linearly depended on the arrival time of the cluster,

$$\gamma_l \propto k_\gamma T_l + \gamma_0. \quad (6)$$

and the mean energy of $l$-th cluster is given by:

$$10\log(\Omega_l) = 10\log(\exp(-T_l/\Gamma)) + M_{cluster}. \quad (7)$$

Assume that we are going to receive signals of the first ray and have achieved perfect synchronization. For the conventional correlation receiver, the template for demodulation is [4]:

$$v(t) = \sum_{j=0}^{N_s-1} p(t - jT_f - C_j^{(1)} T_c - T_1 - \tau_{1,1}). \quad (8)$$

The signal to interference plus noise (SINR) can be expressed as:

$$\text{SINR} = \frac{E_b}{\sigma_n^2 + \sigma_{IASI}^2 + \sigma_{ISI}^2 + \sigma_{MUI}^2}. \quad (9)$$

where, $E_b$ represents the energy for the desired signal, and $\sigma_n^2, \sigma_{IASI}^2, \sigma_{ISI}^2, \sigma_{MUI}^2$ denote the variance for addictive white Gaussian noise, intra-symbol interference, inter-symbol interference and multiuser interference, respectively.

For BPSK system, the BER expression is given by

$$\text{BER} = \frac{1}{2} \text{erfc}\left(\sqrt{\frac{\text{SINR}}{2}}\right). \quad (10)$$

The energy for desired signal $E_b$ and the variance for white Gaussian noise $\sigma_n^2$ are expressed as:

$$\begin{aligned} E_b &= E(Z_u)^2 = \Omega_0 N_s^2. \\ \sigma_n^2 &= E(Z_n)^2 = N_s N_0 / 2. \end{aligned} \quad (11)$$

where, $\Omega_0$ is the energy of the first ray in the first cluster, which can be expressed as:

$$\Omega_0 = \frac{1}{\gamma_0[(1-\beta)\lambda_1 + \beta\lambda_2 + 1]}. \quad (12)$$

The variance for intra-symbol interference $\sigma_{IASI}^2$ is given by

$$\sigma_{IASI}^2 = N_s^2 \sum_{k=2}^{K} \int_0^{T_m} \Omega_0 \exp(-y/\gamma) f_p(y) R^2(y) dy. \quad (13)$$

where, $y$ denotes $\tau_{k,l}$.

And the variance for inter-symbol interference (ISI) $\sigma_{ISI}^2$ is given by

$$\sigma_{ISI}^2 = E(Z_{ISI})^2 = N_s^2 \sum_{k=2}^{K} \int_{-T_m}^{T_m} \Omega_\Sigma \, e^{-y/\gamma} f_p(y) R^2(y) dy. \quad (14)$$

where, $\Omega_\Sigma$ can be expressed as

$$\begin{aligned} \Omega_\Sigma &= \sum_{s=1}^{N_I N_s - 1} E[\Omega_s] \\ &= \frac{1}{2T_s} \sum_{l=1}^{L} \sum_{s=1}^{N_I N_s - 1} \int_{-T_s}^{T_s} \int_0^{sT_f + \tau_{code}} \int_{sT_f + \tau_{code}}^{\tau_{\max}} \Omega_0 \, e^{-T_l/\Gamma} \, e^{-(sT_f + \tau_{code} - T_l)/\gamma} \\ &\quad \times f_c(T_l) f_c(T_{l+1}) dT_l dT_{l+1} d\tau_{code}. \end{aligned} \quad (15)$$

where, $\Omega_s$ represents the $s$-th interfering pulse's energy.

Suppose that there exists $N_u + 1$ users in the system, the variance for multiuser interference $\sigma_{MUI}^2$ is

$$\sigma_{MUI}^2 = E(Z_{MUI})^2$$
$$= F(\omega_0) R_b N_s^2 N_u \sum_{k=2}^{K} \int_{-T_f/2}^{T_f/2} \int_{-z}^{T_m - z} \Omega_0 \, e^{-y/\gamma} f_p(y) R^2(y+z) dy dz$$
$$+ F(\omega_0) R_b N_s^2 N_u \sum_{k=2}^{K} \int_{-T_f/2}^{T_f/2} \int_{-z}^{T_m - z} \Omega_\Sigma \, e^{-y/\gamma} f_p(y) R^2(y+z) dy dz.$$
(16)

where, $y$ represents $\tau_{k,l}$ and $z$ represents $t_u$.

### III. PARTIAL TEMPLATE RECEIVER

For conventional correlation receiver, the template for demodulation is expressed by (8). However, from the simulation results we figure out that such template cannot achieve a desired performance in dense multipath environments such as IEEE 802.15.4a indoor office environment. In dense multipath environment, several adjacent multipaths may overlap with each other and result in intra-symbol interference, which makes the correlation receiver cannot demodulate signals effectively. In this case, we propose a partial template correlation receiver to mitigate intra-symbol interference. The partial template originates from conventional correlation receiver template, which can be obtained by setting the right part of the conventional template to be zero while keep the left part as it is. For instance, if we adopt the second order derivation of Gaussian pulse to transmit information and $N_s$ is chosen to be 1, the conventional template and partial template is shown in Fig. 1 (a) and (b), respectively.

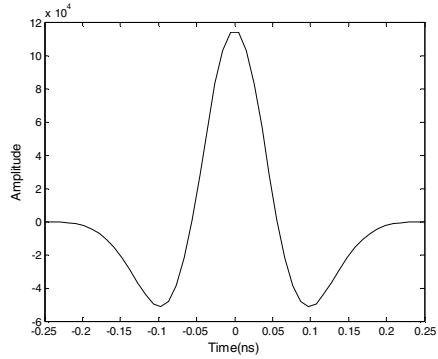

(a) Conventional template

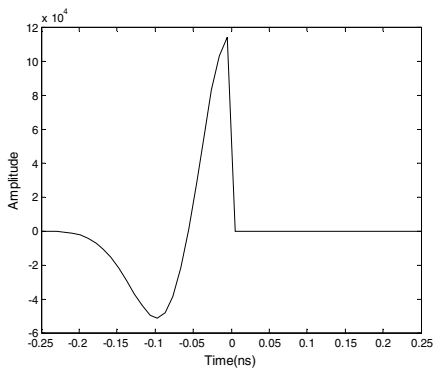

(b) Partial template

Figure 1. The conventional template and partial template

In the meantime, the SINR formulation for partial template receiver is given by

$$\text{SINR}' = \frac{E_b'}{\sigma_n'^2 + \sigma_{IASI}'^2 + \sigma_{ISI}'^2 + \sigma_{MUI}'^2}. \quad (17)$$

where $E_b'$ corresponds to new energy for the desired signal, and $\sigma_n'^2, \sigma_{IASI}'^2, \sigma_{ISI}'^2, \sigma_{MUI}'^2$ represent the new variance for white Gaussian noise, intra-symbol interference, inter-symbol interference and multiuser interference, respectively. The expressions for $E_b', \sigma_n'^2, \sigma_{IASI}'^2, \sigma_{ISI}'^2, \sigma_{MUI}'^2$ are similar to the conventional template receiver discussed in section II, which can be derived by changing the conventional temlplate $v(t)$ to partial template $v'(t)$.

### IV. PERFORMANCE COMPARSION BETWEEN PARTIAL TEMPLATE RECEIVER AND CONVENTIONAL TEMPLATE RECEIVER

In Fig. 2, the second order derivation of Gaussian pulse waveform with time duration $T_m = 0.5\text{ns}$ is adopted in the simulation and 10dB bandwidth is $5.6\text{GHz}$. Without loss of generality, we consider $N_s$ equals 1. Moreover, we set the time hopping width $T_c$ equals to $T_m$ and the number of time hopping $N_h$ is chosen to be 16. Supposing that there exists 4 users in the system with data transmission rate of 15Mbps for each user. We do simulation in indoor office LOS environment defined by IEEE802.15.4a.

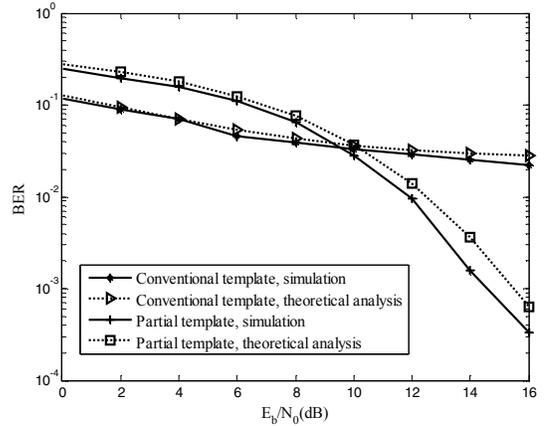

Figure 2. Simulation results for conventional template receiver and partial template receiver

As is shown in Fig.2, partial template receiver demonstrates a better BER performance compared with conventional template receiver. During low $E_b/N_0$ values, the partial template receiver shows a worse BER performance than conventional template receiver, while for the high $E_b/N_0$ values ranges, it has a much better BER performance. More importantly, as the increase of $E_b/N_0$, the partial template receiver shows a more obvious BER advantage over the conventional template receiver.

According to equation (17), when using partial template to demodulate signals, the values of desired signal energy, noise and all types of interferences will decrease. However, during high $E_b/N_0$ values, $\sigma_{IASI}'^2$ has a more significant decrease than the other four decision variables. Therefore, the signal to interference plus noise ratio (SINR) is enhanced indirectly during high $E_b/N_0$ values. Nevertheless, when $E_b/N_0$ is low, the addictive white Gaussian noise becomes the major factor of SINR. Thus, SINR decreases during low $E_b/N_0$ ranges.

Fig. 3 and Fig. 4 present the BER performance comparison between partial template receiver and conventional template receiver in residential LOS environment and indoor office LOS environments, respectively. The parameters for the two channel models are shown in Table 1. Furthermore, different waveforms (the first, third, fifth order derivation of Gaussian pulse) are taken into consideration in the figures.

TABLE I. IEEE 802.15.4A CHANNEL PARAMETERS

|   | *Residential LOS environment* | *Indoor office LOS environment* |
|---|---|---|
| $\bar{L}$ | 3 | 5.4 |
| $\Lambda$ [1/ns] | 0.047 | 0.016 |
| $\lambda_1, \lambda_2,$ [1/ns] $\beta$ | 1.54,0.15,0.095 | 0.19,2.97,0.0184 |
| $\Gamma$ [ns] | 22.61 | 14.6 |
| $\gamma_0$ [ns] | 12.53 | 6.4 |

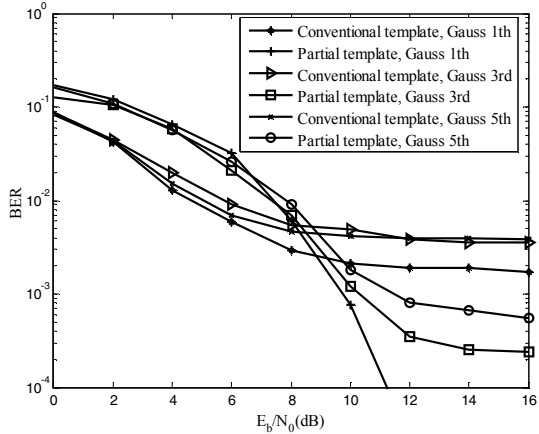

Figure3. BER performance comparison between partial template receiver and conventional template receiver in residential LOS environment

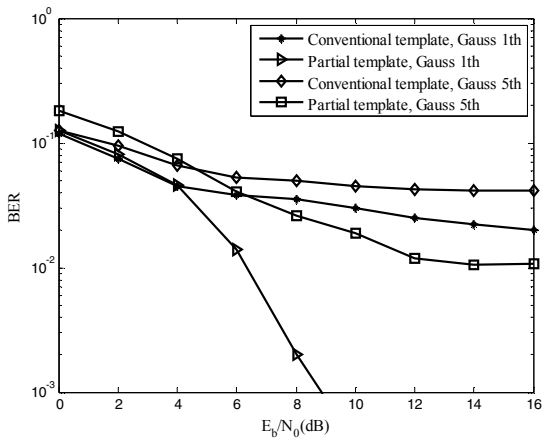

Figure4. BER performance comparison between partial template receiver and whole template receiver in indoor office LOS environments

From Fig. 3 and Fig. 4, we notice that for different channel models and waveforms, the partial template receiver can improve system BER performance in contrast with conventional template receiver during high $E_b/N_0$ values. However, the improvement of system performance varies with different waveforms. It turns out that the first and third order of Gaussian pulse demonstrate a more obvious improvement in BER performance than the fifth derivative of Gaussian pulse.

V. CONCLUSION

In this paper a partial template based receiver for UWB communication systems is introduced. Simulation results show that it can improve system performance during high $E_b/N_0$ values in contrast with conventional template receiver. Moreover, the proposed receiver is a generic receiver, which can be employed in different channel models and different waveforms. Nevertheless, it should be noted that for the low $E_b/N_0$ values, the partial template receiver cannot present an improvement on system performance.